\documentclass{eas}
\usepackage{graphicx}
\def\deg{\ifmmode^\circ\else$^\circ$\fi}
\begin{document}

\title {Relativistic Phenomena in Active Galactic Nuclei}

\author{M.H. Cohen}\address{California Institute of Technology, Pasadena,
CA, 91125}
\author{M.L. Lister}\address{Department of Physics, Purdue University,
   West Lafayette, IN 47907}
\author{R.C. Vermeulen}\address{ASTRON, P.O. Box 2, 7990 AA Dwingeloo,
   The Netherlands}

\begin{abstract} The idea that the radio jets in AGN contain material
in relativistic motion is supported by many lines of observational
evidence, including morphology, brightness temperature estimated
with interferometers and with intrinsic variations, interstellar
scintillations, X-rays, and superluminal motion. These are largely
independent, and taken together make an irrefutable case for relativistic
motion.

\end{abstract}

\maketitle

\section{Introduction}

Scientists were surprised in the early 1950s, when it became clear that
the ``radio stars" in fact were extragalactic and hence very energetic,
and the general idea of a calm orderly Universe had to be questioned. In
later years the discovery of black holes, quasars, pulsars, and other
exotic objects changed the view completely, and brought us to the {\it
violent} Universe as we understand it today.

In a prescient article, Rees (1966) suggested bulk relativistic motion as
an explanation for the rapid variations in flux density seen in quasars;
and he also predicted the phenomenon of superluminal apparent transverse
velocity, which was discovered five years later. At that point most people
believed that cosmic jets contained relativistic jets of plasma, but what
process could collimate and accelerate whole solar masses of material
to relativistic speeds? The current paradigm is that a giant black hole
lies at the center of a galaxy, and the process of accretion combined
with a magnetic field generates the bulk relativistic flow. The details
of this scenario are not well understood, but gravitational accretion is
the only known means for generating the required energy per unit volume.

In this paper we review the radio observations that provide evidence
for relativistic motion.  The plan of the paper is to successively
discuss frequency shift, morphology and Doppler boosting, brightness
temperature measured with interferometers and estimated with intrinsic
variability, interstellar scintillations, inverse-Compton X-rays, and
superluminal motion.  Individually, and especially taken together, these
make a convincing case for relativistic motion in the radio jets. This
conclusion is independent of any specific cosmology.

Relativistic motion implies that the observed radiation patterns are
narrow, that Doppler boosting increases the observed flux density,
and that the observed time scale is shrunk below that in the frame of
the AGN. These affect the observable quantities; namely, frequency,
luminosity, apparent transverse speed, and variability; and there should 
be statistical relations among them.  We briefly touch on this in \S 4.

A number of earlier papers have discussed the different types of
observational data bearing on relativistic motion. We mention only a few
of them here: \cite{Gh93} compared X-ray and VLBI data to derive Doppler
factors and compared them with other beaming indicators; Vermeulen and
Cohen (1994) assembled all the existing data on superluminal sources and
analyzed several models involving the distributions of Lorentz factor and
of the ratio of pattern to beam velocity; Lister and Marscher (1997) made
Monte Carlo calculations of distributions of the measureable parameters,
and of their correlations; and \cite{LV99} estimated Lorentz factors and
also the intrinsic temperature in the synchrotron sources.  These past
surveys mainly used inhomogeneous data culled from the literature. However,
two large surveys on the VLBA, at 5 MHz \cite{VB03} and 15 GHz (Kellermann
\etal~2004, hereafter K04) are now bearing fruit, and analyses with
these superior data sets should soon give much better results.

\section{Formulae}

Here we list some formulae, for later convenience. Assume a relativistic
luminous plasma blob with velocity $\beta = v/c$ and Lorentz factor $\gamma =
(1-\beta^2)^{-{1\over 2}}$. If it is moving at angle $\theta$ to the
line-of-sight (LOS) then its Doppler factor is

$${\delta =} {1\over{\gamma (1-\beta cos{\theta})}}  \eqno(1) $$   

\noindent The observed frequency is $\nu = \delta \nu^*,$ where the
superscript * refers to the coordinate frame of the moving plasma.
Non-starred quantities are in the frame comoving with the AGN.  The
brightness temperature has the same transformation law as the frequency:

$$T_b=\delta T_b^*. \eqno(2) $$

\noindent The observed flux from a point source is

$$S_\nu = \delta^3 S^*_{\nu^*}, \eqno(3) $$

\noindent while for a jet the exponent is 2, and in general the exponent
can be between 2 and 3 \cite{LB85}.

If a source component has proper motion $\mu$ then it has an apparent
transverse speed $\beta_{\rm app}=\mu \rm{D}(1+\it{z)/c}$, where D is
the metric distance to the source. In terms of the source geometry,

$$ \beta_{\rm app} = {\beta sin{\theta} \over {1-\beta cos{\theta}}} \eqno(4)$$

\noindent Graphs of this relation are in Figure~\ref{figthree}.

\section{The Observations}

\subsection{Frequency Shift}

Several galactic sources show radio jets that are analogous to those seen
in the high-luminosity extragalactic sources.  SS 433 contains a jet with
radio-bright components which move out from the core, and the jet changes
orientation in a way that is consistent with precession.  This source
also exhibits frequency-shifted Balmer lines that, when tracked over
time, show conclusively that the speed of the line-emitting clouds is
$\beta = 0.26$ \cite{M80}.  This value is not relativistic, but it is
well above values otherwise associated with stars, including supernovae
shock waves. If SS 433, a compact stellar object, can generate a jet
with $\beta = 0.26$, then we might expect that an extragalactic object
a billion times more luminous can produce a jet that is relativistic.

\subsection{Morphology and Doppler Boosting}

Early imaging showed that many radio galaxies contain lobes more-or-less
symmetrically located across the galaxy, and some objects also contain a
bright point source at the center. A good deal of speculation went into
the question of how these lobes were formed and energized. They were
hundreds or even thousands of kiloparsecs from the center, but their
synchrotron lifetimes were only a million years. If they were ejected
from the galaxy it must have been at relativistic speed,
otherwise there would be a lifetime problem. Alternately, there might
be some kind of continuous beams that kept the lobes energized (eg Ryle
and Longair 1967).

VLA images then showed that the beams did exist, and that there could
be continuous feeding of the lobes. However, in most cases the beams
were strongly asymmetric; one side typically was much stronger than the
other.  Another clue came from VLBI
high-resolution studies. Many compact sources were linear but strongly
non-symmetric; a bright ``core" was at one end of a jet that had a
spectral gradient such that the core had the flattest spectrum. See
K04 for many examples. An important discovery was that the
inner jet nearly always pointed in the direction of the stronger of the
two outer jets.  How could there be one-sided inner structure with
two-sided outer structure? Most astronomers accepted the idea that the
jets contained relativistic beams, and Doppler boosting enhanced the
forward beam relative to the backward beam.

In 1987 the ``Laing-Garington effect" was discovered \cite{GL88}; this was
essentially regarded as a proof of the relativity hypothesis.  The two
outer radio lobes of a radio source typically have different amounts of
Faraday rotation, and the lobe with the weaker Faraday effect is nearly
always on the jet side.  It must be in front, because on average its
radiation goes through less plasma than for the one in back, the non-jet
lobe. The visible jet then is the one in front, and the only convincing
reason that has been offered for this is Doppler boosting.

The front-to-back ratio of the jets, $r$, is typically greater than 10.
This does not call for a very high value of $\gamma$, although 3C 273 is
exceptionally one-sided, with $r>5500$ \cite{D85}. The compact inner
jets often have $r$ = several thousand, which requires $\delta \ge 7$;
i.e. $\gamma \ge 3.5$ and $\theta \le 8.2\deg$. However, these are the
extreme values of $\gamma$ and $\theta$, and more likely values are
$\gamma \sim 7$ and $\theta \sim 6\deg$. These calculations assume that
the jet and counterjet have the same intrinsic luminosity.

Note that the measured high values of $r$ imply small values of $\theta$.
{\it The strongly one-sided objects must be aimed close to the LOS.} What
of the mis-directed ones, those more numerous objects not close to the
LOS? They are not Doppler-boosted; and in fact most will be deboosted
and hence weak. The mis-directed objects are generally thought to be
radio galaxies, and indeed these typically have weak cores \cite{U95}.
The strong, compact, one-sided, flat-spectrum sources often also show
superluminal motion, which also requires an orientation close to the LOS.
(see \S3.6). Given the number of sources that are known to be highly
aligned with the LOS, the parent population is likely to be very large
\cite{LM}. This topic, however, is beyond the scope of this review.

\subsection{Interferometric Brightness Temperature}

A relativistic plasma will produce synchrotron radiation at radio
wavelengths and also X-rays, from the inverse-Compton process. For
reasonable values of the magnetic field the maximum intrinsic temperature
that the plasma can have, if it is in dynamic equilibrium, is 
$\rm T_o \sim 10^{12}$ K, because at a higher temperature second order
scattering becomes important and rapidly quenchs the plasma \cite{KPT}.
Somewhat similar arguments have been made by Singal (1986) and by
\cite{SK94} who derive $3\times 10^{11}$ K and $2\times10^{11}$ K,
respectively, as upper limits.  However, Readhead (1994) has argued
that because at $10^{12}$ K the magnetic and kinetic energies in the
plasma are very far out of balance, a lower equilibrium temperature,
$\rm T_{eq} \sim 5\times 10^{10}$ K, is more likely to be found.

\begin{figure}
  \includegraphics[width=12cm]{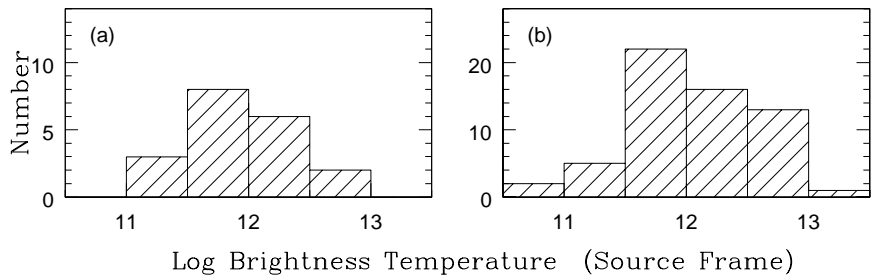}
  \vskip -8 truecm
  \caption{Distributions of interferometric brightness temperature (a)
     measured with the VLBA at 15 GHz, N=19 \cite{Z02}, and (b) measured with
     VSOP at 5 GHz, N=59 \cite{H00}. Note the different vertical scales 
     in (a) and (b).}
\label{figone}
\end{figure}

Measurements made with VLBI show that the temperature can be higher
than these values. Figure~\ref{figone}a shows the distribution of
$\rm{T_b}$ measured with the VLBA at 15 GHz \cite{Z02}, and
Figure~\ref{figone}b shows the distribution measured with the space
VLBI program VSOP at 5 MHz \cite{H00}.  For both the VLBA and VSOP
results, some of the measured values are lower limits.  The two
distributions are similar and show that some sources have $\rm{T_b}$ an
order of magnitude and more above the SSC and other limits and
estimates for $\rm{T_b}$ described above.  Theoretical ways of
exceeding the limits have been proposed, including a non-equilibrium
plasma (Slysh 1992), and coherent radiation processes (Benford and
Lesch 1998).  However, we prefer (with most others) to invoke Eq (2)
and explain the discrepancy with relativistic motion. We expect that
most sources will have $\rm{T_o}\sim 10^{11}$ K in accord with the
estimates of Readhead and others, although we also think there might be
a distribution of $\rm{T_o}$ with a small number of sources approaching
$10^{12}$.  In any event, it is clear that many sources must contain
relativistic beams with $\delta = 5-10$ and more.

Note that this discussion and the one based on morphology are independent
and complementary. They both require Doppler factors up to $\sim 10$.

\subsection{Variability Brightness Temperature}

\subsubsection{Intrinsic Variablity}

\begin{figure}
  \vskip -2 truecm
  \includegraphics[width=12cm]{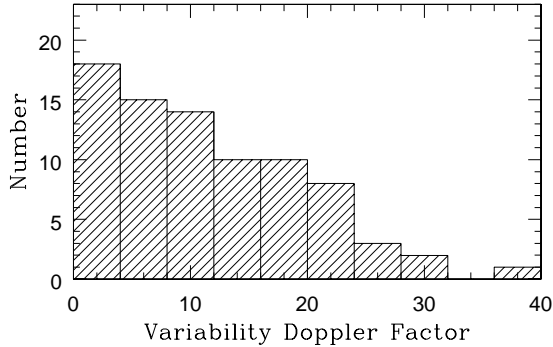}
  \vskip -4 truecm
  \caption{Distribution of Variability Doppler Factor from L\"ahteenm\"aki
     and Valtaoja (1999), with their cosmology corrected to the values used
     in this paper. N=81}
\label{figtwo}
\end{figure}

Many AGN show rapid variability, with time scale $\tau$ as short as a few
weeks. With the assumption that $\tau$ is limited by the light-crossing
time, and with the aid of a model, $\tau$ can be turned into an angular
diameter. The ``variability brightness temperature" T$_{\rm var}$ can then
be found from the flux in the outburst. However, Eq (2) is not valid for
converting these T$_{\rm var}$ to intrinsic values, $\rm T_o$, because
the transformation of $\tau$ introduces another factor of $\delta$, and
$\tau$ enters twice in calculating the solid angle. Hence, in this case
(2) is replaced with

$$\rm{T_{var}=\delta^3 T_o},  \eqno(5)$$                                    

\noindent where $\rm{T_o}$ is the intrinsic temperature in the plasma.
With the assumption of a particular value for $\rm{T_o}$, Eq (5) can be
used to calculate $\rm \delta_{var}$; the cube root makes the estimate of
$\rm \delta_{var}$ somewhat insensitive to the model used in deriving the angular
size.  \cite{LV99} have used their flux monitoring data at 22 and 37
GHz to calculate $\rm{T_{var}}$ and $\rm \delta_{var}$, for $\rm{T_o} = 5\times
10^{10}$ K.  Their histogram of $\rm \delta_{var}$ is in Figure~\ref{figtwo}.
Their values have been converted to the cosmology we use in this paper,
$\rm{H_o=70~km~sec^{-1} ~Mpc^{-1}, ~\Omega_m = 0.3, ~\Omega_\Lambda =
0.7}$.  The distribution of $\rm \delta_{var}$ decreases steadily with
$\rm \delta_{var}$ up to $\rm \delta_{var} \sim 30$.  The model used in
these calculations is an optically thick sphere of radius $c\tau$.

If the interferometric data in Figure~\ref{figone} are converted to
$\delta$ with the same intrinsic temperature as used for the
variability, $\rm{T_o} = 5\times 10^{10}$ K, the $\delta$ are larger
than the ones in Figure~\ref{figtwo}, and a higher temperature, around
$\rm{T_o} = 10^{11}$ K, is needed to bring them into rough agreement. A
detailed comparison cannot be made because the selection criteria for
the two data sets are very different; and also because the
interferometric data contain a number of lower limits on $\rm T_b$.

\subsubsection{Interstellar Scintillations}

Some compact sources show rapid variability due to the passage of
the Earth through the irregular diffraction pattern produced by the
interstellar medium.  This interstellar scintillation can be as fast as
a few hours, giving intraday variability, or IDV.  Scattering theory
gives an estimate of the distance to the screen, and the angular size
of the background source.  Thus the brightness temperature can be
estimated; the method is analgous to VLBI, but with a longer baseline.
Analysis of the IDV cases shows that the screen distance is only 10s
of pc, and the angular scale is 10s of micro-arcsec \cite{DTDB}. The
best-determined case appears to be PKS 0405--385 \cite{R02}, which has
$\rm T_b \sim 2\times 10^{13}$, and $\delta\sim 75$ (for $T_o = 3\times
10^{11}$, the value used by Rickett \etal.  The value $\delta \sim 75$,
while high, is not unprecedented, because the generally-accepted theory
for $\gamma$-ray bursts requires jets with $\delta \sim 100$ or more.
There are many AGN jets with $\delta$ up to about 30, (Fig~\ref{figtwo})
and a few with $\delta$ near 100, but none have been found in-between.
\cite{LV99} state that, if there were sources with variability time
scale $\tau \sim 1$ week at 22 and 37 GHz, they would have been found.
However, in view of the selection biases in all these studies, we think
it premature to generalize these statistics.

\begin{figure}
  \hskip 2.6 truecm
  \includegraphics[width=8cm]{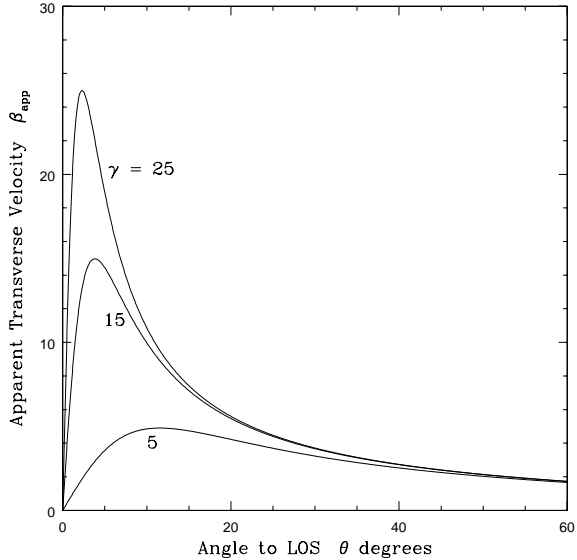}
  \caption{Apparent transverse velocity ${\rm \beta_{app}}$ vs $\theta$,
     the angle to the line-of-sight. Note that every ${\rm \beta_{app}>1}$
     has a maximum possible angle, 
     $\rm \sin\theta_{max} = 2\beta_{app} / (1+\beta_{app}^2)$ 
     and a minimum Lorentz factor $\rm \gamma_{min} = (\beta_{app}^2+1)^{1/2}$}
\label{figthree}
\end{figure}

\subsection{Inverse-Compton X-rays}

A synchrotron source in dynamic equilibrium must radiate X-rays via the
inverse-Compton process, but in some cases the measured X-rays are much
lower than the value expected from the radio flux density. The accepted
explanation for this is relativistic motion, and $\delta$ can be estimated
if the appropriate radio and X-ray measurements are made \cite{M83}.
The calculation is strongly dependent on the radio diameter, which has
to be measured with VLBI. Thus this method is closely allied to the VLBI
temperature method (\S3.3). The X-rays should be measured simultaneously
with the radio, and the radio observations should be made at the peak of
the synchrotron spectrum. These conditions are seldom met.  Nevertheless,
calculations made from literature surveys (eg Ghisellini \etal ~1993)
show that many flat-spectrum sources appear to require $\delta \sim
5-10$. \cite{B99}, however, show that the errors typically are fairly
large, and only a few of the sources in their sample actually require
$\delta > 1$. The best case appears to be 3C 345; Unwin \etal ~(1994,
1997) show that in this source $\delta$ increases from $\sim 5$ to $\ge
10$ as the radiating component moves out from the core. (These values
become somewhat larger if the the cosmology used in this review is used.)

\subsection{Superluminal Motion}

A local observer sees the radiation from a relativistic jet contracted
in time, provided $\delta > 1$, and the $apparent$ transverse speed,
$\beta_{\rm app}$, is greater than the speed of light, in some range
of $\theta$ (Eq. 4).  Figure~\ref{figthree} shows $\beta_{\rm app}$ vs
$\theta$ for several values of $\gamma$; for $\gamma^2 >> 1, ~\beta_{\rm
app}$ is a maximum at $\theta \approx 1/\gamma$, and $\beta_{\rm app,max}
\approx \gamma$. Conversely, for a given $\beta_{\rm app}$, we have $\rm
\gamma_{min} \approx \beta_{app}$.

Early VLBI observations showed a
number of sources that displayed this superluminal effect
\cite{VC}. More recently, systematic large-scale surveys have produced
more than 100 well-defined superluminal sources. In this section we
show some results from a survey on the VLBA, and then discuss
alternatives to the relativistic beam for explaining the observations.

\begin{figure}
  \vskip -1 truecm
  \includegraphics[width=12cm]{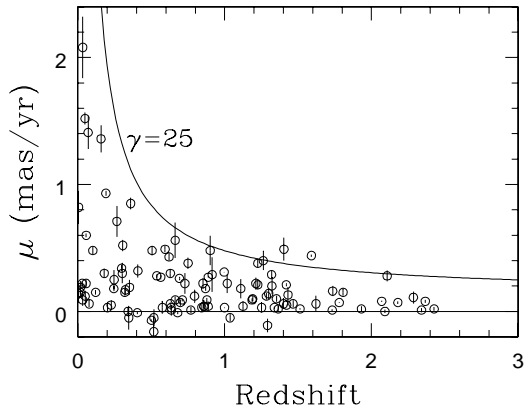}
  \vskip -5 truecm
  \caption{Proper motion vs redshift ($\mu z$ diagram) for the fastest
     ``Excellent" or ``Good" component for 110 sources in the 15-GHz VLBA
     survey.  The line $\gamma=25$ is the locus of points where the minimum
     value of $\gamma$ is 25; a source on the line can have a higher, but
     not a lower, value of $\gamma$.}
\label{figfour}
\end{figure}

\begin{figure}
  \vskip -1 truecm
  \includegraphics[width=12cm]{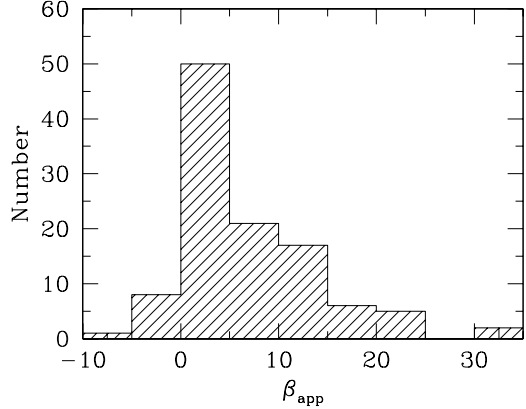}
  \vskip -5 truecm
  \caption{Histogram of the apparent transverse velocity, 
     $\rm \beta_{app}$, for the 110 sources in
     Figure~\ref{figfour}. The median value of $\rm \beta_{app}$ is
     4.2. This also is a histogram of the minimum possible value of $\gamma$.}
\label{figfive}
\end{figure}

Figure~\ref{figfour} shows the $\mu z$ diagram for the fastest components
that are rated E (excellent) or G (good) for 110 sources in the 15-GHz
VLBA survey (K04). The solid line, marked $\gamma = 25$, shows the
points $(\mu, z)$ for which $\gamma \ge 25$.  Hence, the points close
to the line have $\gamma \approx 25$ or greater, and sources below the
line can have correspondingly lower values of $\gamma$. This plot shows
directly that the measured patterns must be moving relativistically.
Figure~\ref{figfive} shows the histogram of $\beta_{\rm app}$ for the
sources in Figure~\ref{figfour}.  This also is a histogram for $\rm
\gamma_{min}$.

We note that the fastest component found in even a rather small sample of
beamed sources with constant $\gamma$ will have $\rm \beta_{app}$ which
closely approaches $\gamma$ \cite{VC}. Hence, we take $\rm \gamma_{max}
= 25$ in our later discussions. In reality the value could be somewhat
larger, but not smaller.

\begin{figure}
  \vskip -1 truecm
  \includegraphics[width=12cm]{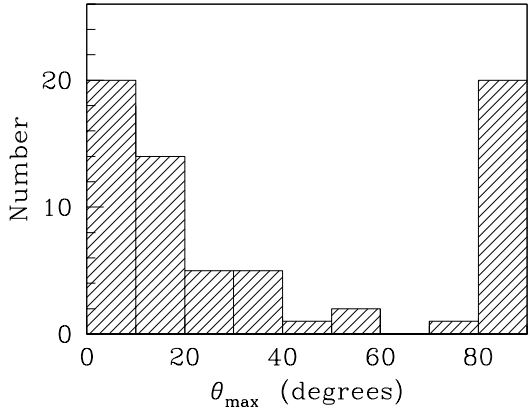}
  \vskip -5 truecm
  \caption{Histogram of $\rm \theta_{max}$, the maximum possible angle to
     the LOS, for the 68 sources in Figure~\ref{figfive} that have
     $\rm S>1.5$ Jy at 15 GHz. The peak near $\rm \theta_{max}=90\deg$
     consists of sources with low or non-significant values of
     $\rm \beta_{app}$ which do not constrain the angle.} 
\label{figsix}
\end{figure}

Any $\rm \beta_{app} > 1$ has a corresponding maximum possible
$\theta$, $\rm \sin\theta_{max} = 2\beta_{app} / (1+\beta_{app}^2)$ (see
Figure~\ref{figthree}).  The histogram of $\rm \theta_{max}$ for the 68
sources with S $>$ 1.5 Jy is in Figure~\ref{figsix}. We picked this flux
density cutoff because it is the limit of the original selection, and
the weaker sources probably introduce a serious bias. These 68 sources,
roughly speaking, approximate a flux-limited sample.  In plotting this
histogram, we set every source with $\rm \beta_{app} < 2\sigma$ or $\rm
\beta_{app} < 1$ to 90\deg; this is responsible for the peak at 90\deg.

In fact, $\rm \theta_{max}$ is highly unlikely to be the actual value
of $\theta$ for a beamed source.  It can be shown that the most
probable viewing angle, $\theta_p$, for a source picked on the basis of
its flux density from a randomly-oriented beamed population, is about
$1/(2\gamma)$ \cite{LM}. Since the minimum $\gamma$ is approximately $\rm
\beta_{app}$, a rough upper limit to $\theta_p$ is $\rm
\theta_{max}/4$.  We conclude that the histogram in
Figure~\ref{figsix} is unlikely to be representative of the real
values of $\theta$; a more probable distribution would be more peaked
at small angles.


Lister and Marscher (1997) show some plots of the distribution of $\rm
\beta_{app}$ and $\theta$ from Monte-Carlo simulations of observing a
complete well-defined sample of relativistically boosted jets.  We have
begun to observe such a sample, in a continuation of the 15 GHz VLBA
survey called the MOJAVE project \cite{L03}.

\subsubsection{Pattern Speed vs Beam Speed}

It is usually assumed that the speed measured for the moving
pattern is also the speed of the beam, but this often may be wrong
\cite{LB85}. Shocks will exist in the jet, and in some cases the moving
blobs will be compressions due to the shocks \cite{HAA85}. In these cases
the pattern speed can be different from the underlying flow velocity.
As an extreme case, consider a standing shock wave set up at a bend in the
jet; the measured speed (zero) gives no information on the actual flow
velocity.  Another possibility is that a beam contains two components,
a narrow fast spine and a slower sheath. In this case the observed pattern
speed might be a combination of the two speeds, although we might see only
the spine, because its Doppler boosting will be higher.  Other effects,
including bending of the beam and evolution in the moving component, can
also give erroneous speeds.  Nevertheless, in most cases the measured
pattern velocity should be similar to the actual beam speed, because
the speeds are roughly compatible with other relativity indicators.
(See \S4.4).

\subsubsection{Alternatives to the Relativistic Beam Model}

Relativistic beams are generally accepted as the explanation for
superluminal motion, but alternatives have been suggested and are still
occasionally referred to. These can be grouped into two main categories.

\noindent 1. Random

Among the 110 sources of the 15-GHz survey, nearly all show expansion or
zero velocity. The two sources with a statistically significant ($2\sigma$)
negative velocity can be reasonably explained with outward helical motion
or with a bent jet, as an alternative to contraction. Further, in most
cases the motions are along a line, and the expansions can persist for up
to a decade in distance from the core. In the two-sided source NGC 1052
\cite{V03}, the speeds are similar on the two sides. These motions
are not random.

\noindent 2. Non-beaming models

The light-echo model (Lynden-Bell 1977) works for supernova remnants
but not for jets in AGN, because the predicted distribution of
velocities does not match the data.  The same problem arises for echoes
in a dipole magnetic field \cite{BM80}.  Ekers \& Liang (1990) produced
an ingenious model of a relativistic excitation wave in a jet, where
the radiating material is stationary; \ie~it is not swept up by the
wave. This suffers from the same problem with the distribution of
velocities.

These models do not have relativistic motion of the radiating material,
so there is no Doppler favoritism. The selected sources must be
oriented randomly relative to the LOS. On the other hand, the
distribution in Figure~\ref{figsix}, based only on the observed values
of $\rm \beta_{app}$ and without explicitly or implicitly assuming
Doppler boosting, shows that the orientations are not random, which
would give a $\sin\theta$ distribution.

This result was quantified by \cite{C90} with an analysis of the 12
strongest sources that were then known.  Each source has a solid angle
associated with $\rm \theta_{max}$ that can be compared with $2\pi$, the
total available, in a manner analagous to a $\rm{V/V_m}$ calculation.
The result was that the probability that the objects were drawn from a
randomly oriented parent population was small unless the Hubble constant
was well above any reasonable value. This calculation can now be repeated
with a much better and larger sample, and with a current value for $\rm
{H_o}$. The result for the fastest component that is rated E or G, for
the 53 quasars that have S(15)$>1.5$ Jy (the original limit of the 15-GHz
survey), is that the probability that they were drawn from a randomly
oriented parent population is $p < 6 \times 10^{-6}$. This is a strongly
biased estimate because (a) a distribution based on a more probable angle,
rather than the maximum angle, would have a smaller probability, and
(b) every source with $\beta_{\rm app} < 2 \sigma$ or $\rm \beta_{app}
< 1$ was set to 90 degrees. Thus the estimate for $p$ is conservative,
and we conclude that the quasars are not oriented at random, but rather
are preferentially close to the LOS. The models without Doppler favoritism
are not viable.

An exception to this conclusion could arise if there were obscuration near
the equatorial plane, so that small values of $\theta$ would be favored.
However, we do not see how this could come about. The dust clouds
invoked for optical obscuration are transparent at radio wavelengths,
and free--free absorption at 15 GHz would require unreasonable densities
and masses of plasma.

\section{Self-Consistency}

The brightness temperature of a source can be determined with
interferometry ($\rm T_b$) and also with flux density variations ($\rm
T_{var}$). Comparisons of the two measures can help determine if the
various assumptions underlying these determinations are valid. If $\rm
T_b$ and $\rm T_{var}$ are measured then the Doppler factor $\delta$ and
the intrinsic temperature ($\rm T_o$) can be calculated with Equations
2 and 5.  L\"ahteenm\"aki \etal ~(1999a) have done this for several
sets of sources, and found that most of the $\rm T_o$ have a range
around $10^{11}$ K. This is rather lower than the SSC limit but close
to the other values discussed in \S3.3. This suggests that, although
the detailed variability model commonly used is not too realistic,
the results none-the-less should have some reliability, and the different
measurements pertain to the same relativistic plasma. As discussed in
\S 3.4.1, the sets of $\delta$ found by the two methods are in rough
agreement with $\rm T_o = 10^{11}$.

Now we consider the relation between the variability Doppler factor
$\rm \delta_{var}$ and the apparent transverse speed $\rm \beta_{app}$,
and assume that the same plasma motion affects both phenomena, through
the shrinkage of the time scale by the forward relativistic motion.
They should be closely connected, but note that high $\rm \delta_{var}$
can go with low $\rm \beta_{app}$, when $\theta < 1/\gamma$.  If $\rm
\delta_{var}$ and $\rm \beta_{app}$ are known, we can calculate $\gamma$
and $\theta$ from Eqs 1 and 4, but there is ambiguity because an assumed
value of $\rm T_o$ is required for $\rm \delta_{var}$.  We test the
paradigm by calculating $\rm \delta_{var}$ with several values of
$\rm T_o$, and comparing the resulting distribution of points $(\rm
\beta_{app}, \delta_{var})$ with a Monte-Carlo simulation.  This should
show whether or not $\rm \delta_{var}$ and $\rm \beta_{app}$ could
be related to each other through Eqs 1 and 4, and it might suggest an
optimum value for $\rm T_o$.

The simulation is of the type described by Lister and Marscher (1997). It
uses a set of beamed sources that are uniformly distributed in space
and isotropically distributed in angle, and have power law distributions
of luminosity and Lorentz factor. For the simulation shown here we use
power law indices of -2.0 and -1.5, respectively.  The calculation draws
sources at random and counts those that are above a flux limit.

\begin{figure}
  \vskip -2.5 truecm
  \includegraphics[width=12cm]{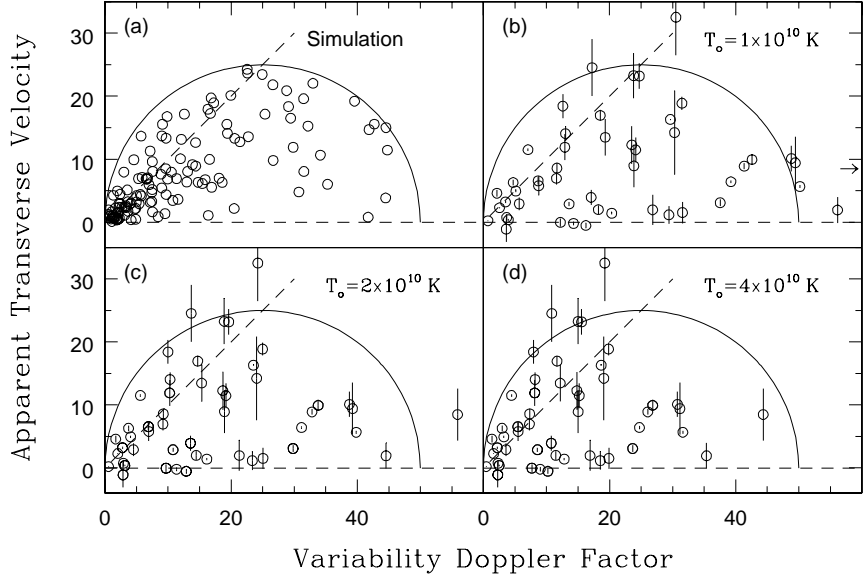}
  \vskip -1 truecm
  \caption{Simulated and measured values of $(\rm \beta_{app},
     \delta_{var})$.  (a) Simulated flux-limited
     sample (N=133) of beamed sources having $\rm \gamma_{max}=25$. The
     solid line is the locus of points  for $\gamma=25$; the dashed line
     $\rm \beta_{app} = \delta_{var}$ marks the ``$1/\gamma$ cone" around the
     LOS (see text).  (b) The points $(\rm \beta_{app}, \delta_{var})$ for
     the fastest component of the sources having both a $\rm \beta_{app}$
     value from the 15 GHz VLBA survey (K04), and a $\rm \delta_{var}$
     value from variability studies at 22 and 37 GHz \cite{LV99} (N=49). The
     $\rm \delta_{var}$ are calculated with intrinsic temperature $\rm T_o =
     1 \times 10^{10}$. (c) as in (b) with $\rm T_o = 2 \times 10^{10}$.
     (d) as in (b) with $\rm T_o = 4 \times 10^{10}$.}
\label{figseven}
\end{figure}

Figure~\ref{figseven} shows $\rm \beta_{app}$ plotted against $\rm
\delta_{var}$ for the 49 sources that have both an ``excellent" or ``good"
value of $\rm \beta_{app}$ (K04) and a value of $\rm \delta_{var}$ from
the 22 and 37 GHz variability studies of \cite{LV99}.  We have eliminated
five sources with speed consistent with zero and located at a sharp bend
in the jet, because we assume that they are connected to stationary shock
waves and are not indicative of the motion.  The Monte-Carlo simulation
is in the top left panel; the other panels represent the observations
with $\rm \delta_{var}$ calculated with different values of $\rm T_o$.
These plots are discussed further in \cite{C03}.  The solid lines in
Figure~\ref{figseven} represent $\gamma = 25$. A sample of sources
with a distribution of $\gamma$ with $\rm \gamma_{max}=25$ will be
entirely below the line. We used $\gamma=25$ following the discussion
of Figure~\ref{figfour}, and it appears to provide a reasonable envelope.

The probability of finding a beamed source selected on the basis of
flux density is maximized well inside the $1/\gamma$ cone \cite{VC};
and 72\% of the sources in the simulation are indeed inside the cone.
In Figures~\ref{figseven}b,c,d the fractions are 86\%, 71\%, and 63\%,
respectively; and on this basis we would select (c) as providing the
best fit to the simulation. But the detailed distribution also suggests
that (c) is preferable to (b) or (d), In (b) there is a cluster of
points outside the $\gamma=25$ circle including one at $\rm
\delta_{var}=70$; however it is hard to judge the importance of this
because the errors on $\rm \delta_{var}$ are large and not
well-understood. The effect becomes exaggerated if $\rm T_o$ is reduced
below $1 \times 10^{10}$. In (d) the points appear to be clustered
rather far to the left, and this is exaggerated if the temperature is
raised. The result is that the speeds measured with the VLBA are
consistent with the variability Doppler factors provided that the
intrinsic temperature is roughly $2\times 10^{10}$ K.

The number of points is small and the errors are large, but in
Figure~\ref{figseven}c, for $\rm T_o=2\times 10^{10}$ K, the distribution
is perhaps matched to the simulation. \cite{C03} discuss the fit.  $\rm
T_o=2\times 10^{10}$ K can be taken as a rough value of the intrinsic
temperature. In fact there surely is a distribution of $\rm T_o$, but
there are not enough data to study that.

\begin{figure}
  \vskip -6 truecm
  \includegraphics[width=12cm]{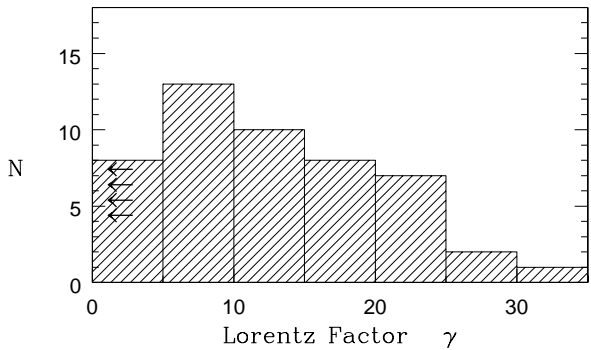}
  \vskip -1 truecm
  \caption{Histogram of Lorentz factor calculated from $\rm \delta_{var}$
     and $\rm \beta_{app}$ with $\rm T_o=2\times 10^{10}$ K.  Arrows indicate
     upper limits.}
\label{figeight}
\end{figure}

Figure~\ref{figeight} shows the distribution of $\gamma$ calculated
from the $\rm (\beta_{app}, \delta_{var})$ points, with $\rm
T_o=2\times 10^{10}$ K. Four sources have formally negative values of
$\rm \beta_{app}$, and in Figure~\ref{figeight} they are plotted
as upper limits at the $2\sigma$ position. The values
above $\gamma=25$ can be seen above the line in
Figure~\ref{figseven}. Two of them are within $2\sigma$ of $\gamma=25$,
and the third, like all the points, has an unknown error connected with
the variability Doppler factor.

The value $\rm T_o \sim 2\times 10^{10}$ K from the $\rm (\beta_{app},
\delta_{var})$ distribution is rather different from the value $\rm T_o
\sim 10^{11}$ K from the comparison of interferometer and variability
Doppler factors. However, these agree within an order of magnitude and
perhaps this is a success, given the many assumptions going into the
analyses, and the large observational errors.

\section{Conclusions}

We have described five independent sets of radio phenomena which show
that there is relativistic motion in the compact jets of AGN. These are
the observations of morphology, interferometry, intrinsic variability,
interstellar scintillations, and superluminal motion. X-ray measurements
provide further confirmation, although these results are dependent on the
VLBI measurements from which the brightness temperatures are determined.
Each of these phenomena has a theoretical difficulty which is readily
alleviated by assuming that there is bulk relativistic motion towards the
observer.  The morphology, interferometry, interstellar scintillations,
and the X-ray method are independent of redshift and hence of cosmology.
Interpretations based on intrinsic variability and on internal proper
motion, however, require a knowledge of the distance to the source. The
variability methods, and the interferometry method, require an assumption
of an intrinsic temperature in the synchrotron cloud.  The distribution
of $\delta$ peaks at low values and decreases to $\delta \sim 30$ The
Lorentz factor $\gamma$ is a more fundamental quantity than $\delta$,
but is harder to estimate.  Values estimated for $\gamma$ range up to
34.

A comparison of $\rm \beta_{app}$ and $\rm \delta_{var}$ shows that
the superluminal speeds and the variability Doppler factors are roughly
consistent with the standard relativistic motion paradigm, provided the
intrinsic temperature in the synchrotron clouds is $\sim 2\times 10^{10}$ K.
This value is below the various theoretical upper limits, and is close to 
Readhead's estimate (1994), based on equipartition between the energy in
the particles and that in the magnetic field.


\begin{thebibliography}{}
\bibitem[(Bahcall \& Milgrom 1980)]{BM80}Bahcall, J.N. \& Milgrom,M. 1980
    ApJ, 236, 24
\bibitem[()]{} Benford, G. \& Lesch, H. 1998, MNRAS, 301, 414
\bibitem[Bloom \etal ~(1999)]{B99} Bloom, S. D. \etal ~1999, ApJS, 122, 1
\bibitem[Cohen (1990)]{C90} Cohen, M. H. 1990, in Parsec-Scale Radio
    Jets, ed. J.A. Zensus \& T.J. Pearson, (Cambridge: Cambridge University
    Press), 317
\bibitem[Cohen \etal ~(2003)]{C03} Cohen, M. H. \etal, 2003 in Radio Astronomy 
   at the Fringe, ASP Conference Series No 300, eds JA Zensus \etal, 177
\bibitem[(Davis \etal~1985)]{D85} Davis, R.J., Muxlow, T.W.B. \&
   Conway 1985, Nature, 318, 343
\bibitem[(Dennett-Thorpe \& de Bruyn 2000)]{DTDB} Dennett-Thorpe, J. \&
   de Bruyn, A. G. 2000, ApJ, 529, L65
\bibitem[()]{} Ekers, R.D. \& Liang, H. 1990, 
    in Proc. Workshop on Parsec-Scale Jets, (Cambridge: CUP), 333
\bibitem[(Garrington \etal ~1988)]{GL88} Garrington, S.T., Leahy J.P.,
    Conway R.G. \& Laing R.A. 1988, Nature, 331, 147
\bibitem[Ghisellini \etal ~(1993)]{Gh93} Ghisellini, G., Padovani, P.,
    Celotti, A. \& Maraschi, L. 1993, ApJ, 407, 65
\bibitem[(Hirabayashi \etal ~2000)]{H00} Hirabayshi, H. \etal ~2000, PASJ, 52, 997
\bibitem[(Hughes \etal ~1985)]{HAA85} Hughes, P. A., Aller, H. D. \& Aller M. F. 
    1985, ApJ, 298, 301
\bibitem[(Hjellming \& Johnston 1981)]{HJ81} Hjellming, R. M.  \& Johnston, K. J.
    1981, ApJ, 246, L141
\bibitem[(Kellermann \& Pauliny-Toth 1969)]{KPT} Kellermann, K.I. \&
    Pauliny-Toth, I.I.K. 1969, ApJ, 155, L71
\bibitem[()]{} Kellermann, K.I. \etal~2004, in press; K04
\bibitem[L\"ahteenm\"aki \& Valtaoja (1999)]{LV99} 
     L\"ahteenm\"aki, A., \&  Valtaoja, E. 1999, ApJ, 521, 493
\bibitem[()]{} L\"ahteenm\"aki, A., Valtaoja, E., \& Wiik, K. 1999a, ApJ, 51l, 112
\bibitem[(Lind \& Blandford 1985)] {LB85} Lind, K. R. \& Blandford, R. D. 
    1985, ApJ, 295, 358
\bibitem[()]{} Lynden-Bell D. 1977, Nature, 270, 396
\bibitem[(Lister 2003)]{L03} Lister, M. 2003, in "Future Directions in 
    High Resolution Astronomy: A Celebration of the 10th Anniversary of the VLBA, 
    eds J. Romney \etal, in press.
\bibitem[(Lister \& Marscher 1997)]{LM} Lister, M. L. \& Marscher, A. P. 1997,
    ApJ, 476, 572
\bibitem[(Marscher 1983)]{M83} Marscher, A.P. 1983, ApJ, 264, 296
\bibitem[(Margon \etal ~1980)]{M80} Margon, B., Grandi, S.A. \& Downs, R. A.
    1980, ApJ, 241, 306
\bibitem[()]{} Readhead, A.C.S. 1994, ApJ, 426, 51
\bibitem[()]{} Rees, M. J. 1966, Nature, 211, 468
\bibitem[(Rickett \etal ~2002)]{R02} Rickett, B. J., Kedziora-Chudczer, L.,
    \& Jauncey, D. L. 2002, ApJ, 581, 103
\bibitem[()]{} Ryle, M. \& Longair, M.S. 1962, MN, 136,123
\bibitem[Sincell \& Krolik (1994)]{SK94} Sincell, M. W. \& Krolik, J. H. 
   1994, ApJ, 430, 550 
\bibitem[{}]{} Singal, A. K. 1986, A\&A, 155, 242 
\bibitem[{}]{} Slysh, V.I. 1992, ApJ, 391, 453
\bibitem[(Unwin \etal ~1994)]{U94} Unwin, S. C., Wehrle, A. E., Urry,
   C. M., Gilmore, D. M., Barton, E. J., Kjerulf, B. C., Zensus, J. A. \&
   Rabaca, C. R. 1994, ApJ, 432, 103
\bibitem[(Unwin \etal ~1997)] {U97} Unwin, S. C, Wehrle, A. E., Lobanov, A. P.,
   Madjeski, G. M., Aller, M. F. \& Aller, H. D. 1997, ApJ, 480, 596
\bibitem[(Urry \& Padovani 1995)]{U95} Urry, C.M. \& 
    Padovani, P. 1995, PASP, 107, 803
\bibitem[(Vermeulen \etal~2003)]{VB03} Vermeulen, R.C., Britzen, S.,
   Taylor, G.B., Pearson, T.J., Readhead, A.C.S., Wilkinson, P.N. \&
   Browne, I.W.A. 2003, in Radio Astronomy at the Fringe, ASP Conference 
   Series No 300, eds JA Zensus \etal, 43 
\bibitem[(Vermeulen 2003)]{V03} Vermeulen, R. C. \etal ~2003, A\&A, 401, 113
\bibitem[(Vermeulen \& Cohen 1994)]{VC} Vermeulen, R.C. \& Cohen, M.H. 
    1994, ApJ, 430, 467
\bibitem[(Zensus \etal ~2002)] {Z02} Zensus, J. A., Ros, E., Kellermann, K. I.
    Cohen, M. H., Vermeulen, R.C. \& Kadler, M. 2002, AJ, 124, 662




\end{thebibliography}
\end{document}